\documentclass[a4paper, 10 pt, conference]{ieeeconf} 
\IEEEoverridecommandlockouts                            
 \pdfoutput=1
\usepackage{float}
\usepackage{caption}
\usepackage{subcaption}
\usepackage{cite}
\usepackage{amsmath, amssymb}
\usepackage{amsfonts}
\usepackage{graphicx}
\usepackage{enumerate}
\usepackage{titlesec}
\usepackage{ifthen}
\usepackage{fancyhdr}
\usepackage[usenames, dvipsnames]{color}
\usepackage{lipsum}
\usepackage{mwe}
\graphicspath{{figures/}}
\usepackage{mathtools}
\usepackage{multicol}
\usepackage{xcolor}
\usepackage{tikz}
\usepackage{pgfplots}
\pgfplotsset{compat=newest}
\usetikzlibrary{patterns}
\usetikzlibrary{decorations.text}
\usepgfplotslibrary{fillbetween}

\definecolor{disagreement}{rgb}{1, 0.5, 0.5}
\definecolor{consensus}{rgb}{0.5, 0.5, 1}

\usepackage{cancel} %To be able to cross out text
\usepackage[normalem]{ulem}
\usepackage{flushend}
\usepackage{comment}
\newtheorem{theorem}{Theorem}%[section]
\newtheorem{corollary}[theorem]{Corollary}
\newtheorem{lemma}[theorem]{Lemma}%[section]
\newtheorem{remark}{Remark}
\newtheorem{proposition}[theorem]{Proposition}%[section]
\newtheorem{definition}{Definition} %[section]
\newtheorem{example}{Example}%[section]

\newboolean{showcomments}
\setboolean{showcomments}{true}

\newcommand{\luka}[1]{\ifthenelse{\boolean{showcomments}}
{\textcolor{black}{(Luka says: #1)}}{}}
\newcommand{\david}[1]{\ifthenelse{\boolean{showcomments}}
{\textcolor{black}{(David says: #1)}}{}}
\newcommand{\emma}[1]{\ifthenelse{\boolean{showcomments}}
{\textcolor{VioletRed}{(Emma says: #1)}}{}}
\newcommand{\edit}[1]{\ifthenelse{\boolean{showcomments}}
{\textcolor{black}{#1}}{}}

\usepackage{currfile}
\usepackage{cite}

\allowdisplaybreaks

\title{\LARGE \bf
Multipolar opinion evolution in biased networks
}

\author{ {Luka Baković, David Ohlin, Giacomo Como and Emma Tegling} 
 \thanks{The authors are with the Department of Automatic Control, Lund University, Lund, Sweden. They are members of the ELLIIT Strategic Research Area at Lund University.   Email: \{{\tt\small{luka.bakovic,david.ohlin, giacomo.como, emma.tegling}\}@control.lth.se}.  }
        \thanks{Giacomo Como is also with the Department of Mathematical Sciences, Politecnico di Torino, Torino, Italy. {\tt\small{giacomo.como@polito.it}}}
        \thanks{This work was partially supported by ELLIIT and the Wallenberg AI, Autonomous Systems and Software Program (WASP) funded by the Knut and Alice Wallenberg Foundation.}
}

\usepackage{url}

\begin{document}

\maketitle
\thispagestyle{empty}
\pagestyle{empty}

%%%%%%%%%%%%%%%%%%%%%%%%%%%%%%%%%%%%%%%%%%%%%%%%%%%%%%%%%%%%%%%%%%%%%%%%%%%%%%%%
\begin{abstract}
Motivated by empirical research on bias and opinion formation, we \textcolor{black}{formulate} a multidimensional nonlinear opinion-dynamical model where agents have individual biases, which are fixed, as well as opinions, which evolve. \textcolor{black}{The dimensions represent competing options, of which each agent has a relative opinion, and are coupled through normalization of the opinion vector.}
This can capture, for example, an individual's relative trust in different media. In special cases including where biases are uniform across agents our model achieves consensus, but in general, behaviors are richer and capture multipolar opinion distributions. We examine general fixed points of the system, as well as special cases such as zero biases toward certain options or partitioned decision sets. Lastly, we demonstrate that our model exhibits polarization when biases are spatially correlated across the network, while, as empirical research suggests, a mixed community can mediate biases.
\end{abstract}

\section{Introduction}

%Social media is becoming increasingly relevant as a news source, as more and more people use it either as a way of aggregating content from different sources directly or gathering information by reading articles shared by their friends. According to \cite{pew2020news}, more than half of Americans "at least sometimes" get news from social media, and for millennials \cite{eu2019barameter} shows that social media is even the primary news source.

Social media is becoming increasingly relevant as a news source, as more and more people use it either as a way of aggregating content from different sources directly or gathering information by reading articles shared by their friends. \textcolor{black}{More than half of Americans get news from social media, and for millennials \cite{eu2019barameter} shows that social media is even the primary news source.}
With the range of news sources available, a key question is how users decide whom to trust. 
While \emph{accuracy} of the news is an important contributing factor to media trust \cite{Schranz2018}, it is also true that perceived accuracy may be subjective. For example, left, center and right leaning news outlets all have their audiences, which would imply three differing viewpoints being accurate at the same time. Indeed, \emph{bias} plays a key role and has long been the subject of social science research. As shown in \cite{Nickerson1998}, individuals have a tendency to gather evidence which is partial to their preconceptions. In other words, we perceive something as accurate if it is in line with our biases and preconceptions, see \cite{Mullainathan2005} for an elaboration. In this paper, we propose a novel opinion-dynamical model that \textcolor{black}{can} capture how preconceptions or biases impact people's evolving trust in the news sources that their social networks expose them~to.

Recent empirical research on social networks shows that they do in fact exhibit phenomena such as polarization \cite{Urman2019} and ideological segregation \cite{GonzlezBailn2023}, meaning that biases seem to play a role in how information is distributed and opinions are formed. 
It is clear, however, that the mapping between bias and opinions is not an identity map. Indeed, as shown in research on contact theory, outlined in \cite{allport1954prejudice}, intergroup contact can act as a mediator for biases. Sufficiently diverse networks can imply bias mediation. This is found in \cite{Bai2020diversity} to be the case -- greater diversity can lead to greater perceptions of similarity. This phenomenon is also captured by our model.

With this empirical context in mind, the contribution of this paper is to propose a novel multidimensional nonlinear opinion-dynamical model where agents \textcolor{black}{have an underlying bias}, as well as opinions, \textcolor{black}{which} evolve \textcolor{black}{as} the information shared by their neighbors in a social network \textcolor{black}{is filtered through the bias}. \edit{More specifically,} agents in our model update their opinion by combining it with an individually weighted (i.e. biased) average of their neighbors' opinions in each dimension. The vector components, that is, the opinions on different issues, are then coupled through a normalization that ensures the opinion vector sums to 1. The opinion vector may, for example, represent an individual's relative trust in, or attention to, various news sources, where an increase in attention to one source will lead to a decrease in another. The nonlinearity allows the model expressivity in the sense that it exhibits different steady states ranging from polarization to consensus, depending on the configuration and initial conditions. For example, introducing spatial correlation between biases leads to a polarization of opinions. In other words, communities in the graph will become echo chambers if given similar biases, while diversity can mediate the biases.

%A representative example is that of media consumption. There, agents' opinion vectors can represent their opinions toward $k$ different media, with their biases being their political orientation. Attention toward the media is limited, as one cannot read all of the sources at once, which is adequately captured by the fact that opinions are necessarily unit vectors in our model.

\subsection{Related work}
Some of the most commonly studied models in opinion dynamics are linear ones which use averaging as principle mechanism for social interaction. Such are the DeGroot model \cite{degroot1974reaching} and the Friedkin-Johnsen model \cite{friedkin1990social, friedkin1999influence}. A detailed analysis of both can be found in \cite{proskurnikov2017tutorial,proskurnikov2018tutorial}. Both models predict convergence to a convex combination of the starting opinions, and are therefore unsuitable for modeling of radicalization, formation of echo chambers or segregation, which real networks do exhibit as shown in \cite{Urman2019} and \cite{GonzlezBailn2023}. Compared to them, our model is more expressive as it only reaches consensus in some specific cases and allows opinions to escape the convex hull of initial opinions. Our model allows like-minded individuals to push each other further toward extreme opinions. It is worth noting, however, that for a certain choice of parameters (uniform bias) our model reduces to the DeGroot model.

Conceptually, our model also relates to the bounded confidence model introduced in \cite{Deffuant2000}. There, agents only interact with agents whose opinions are close enough to theirs\textcolor{black}{, meaning that the network itself evolves with the opinions.} Whilst successfully capturing segregation of opinions, this model still \textcolor{black}{converges to} convex combinations of starting opinions. Multidimensional versions of \cite{Deffuant2000} also exist, for example in~\cite{Lorenz}. \textcolor{black}{A related multidimensional model is proposed by the simulation study \cite{Schweighofer2020}, where the interaction between agents is determined by their agreement across the opinion vector, also accounting for negative influence. This is in contrast to our model, where agents with different opinions can still interact if their biases permit it. A more intricate coupling mechanism motivated by epidemical modeling is proposed by~\cite{Peng2021-cr}. There, the vaccination network can be compared to our bias and the epidemic to the opinion. Both networks then evolve dynamically. whereas we fix the bias and consider opinion dynamics only.} 
%Their vaccination network can be viewed as a bias, and the epidemical as an opinion. The difference is that both networks dynamically evolve, whereas we fix bias and look purely at opinion dynamics.}

In certain aspects, our model also relates to \cite{Parsegov2017}, wherein a linear operator is applied to the "neighborhood" term of the averaging dynamics, thus coupling the opinion vector. The nonlinearity of our model, however, captures other behaviors. The nonlinear opinion dynamics model presented in \cite{Bizyaeva2023} has recently received a lot of attention. Our models exhibit similarities such as the ability for multi-option opinion formation and parameters mimicking preference or bias of agents toward certain issues. Both models include some simpler models as special cases. Our models differ in domain choice and type of nonlinearity, as well as the underlying empirical motivation. \textcolor{black}{As opinions evolve, agents in our model can stop adapting to certain neighbors. This is not unlike the self-appraisal mechanism of~\cite{Jia2015}.}

\subsection{Contributions and paper outline}
This paper proposes a novel multidimensional nonlinear opinion-dynamical model, motivated by the body of empirical work on bias and preconceptions. It provides an initial analysis of its behaviors that depend on the biases, the opinions as well as the social network structure. We provide a characterization of fixed points of the system and give sufficient conditions for the special case where agents reach a consensus. We also present a Lyapunov argument for the case where one set of options is strictly preferred by all agents. We illustrate the dynamics through numerical examples, which in particular demonstrate emergence of polarization in cases where biases are spatially correlated.

The model is introduced in Section~\ref{modelintro}. Section~\ref{anaylsis} contains analytical results on the system's fixed points, which are then illustrated through numerical examples in Section~\ref{section:numerical}. We conclude with directions for future work in Section~\ref{sec:conclusion}.
%The main contribution of this paper is the introduction of a novel multidimensional nonlinear opinion-dynamical model. It is motivated by a sizable body of work in social science studying bias and preconceptions and aims to fill in the gap of expressive models with a plethora of convergence modes dependent on network structure, agent biases and opinions. 

% In Section~\ref{modelintro} we present the model itself, draw comparisons to preexisting models and prove that our model includes DeGroot averaging as a special case.

% Section~\ref{anaylsis} contains analytical results on fixed points. We provide a set of trivial fixed points and provide an example system which reaches them. After introducing a general proposition on the behavior of agents at fixed points, we introduce corollaries for cases where agents end up on the edges of the domain. Finally, we present a Lyapunov function argument for the convergence of the system in cases where one set of options is strictly preferred by all agents. 

% Section~\ref{section:numerical} provides two examples. The first one serves to provide intuition on the two agent case, showing the three modes of convergence depending on the biases and initial opinions on the agent. The second one is a larger simulation showcasing polarization under spatial correlation of biases, a complex behavior which manifests when dynamics take place over a Watts-Strogatz network.

\section{The Model} \label{modelintro}

\subsection{Notation and Preliminaries}
Bold fonts indicate vectors, capital letters indicate matrices. Let $\mathcal G = \{ \mathcal V, \mathcal E \}$ \textcolor{black}{be a finite simple graph, where $\mathcal V$ is the set of nodes and $\mathcal E$ undirected edges. Simple graphs are those with no multi-edges or self-loops. A simple graph is connected if there exists a path between every pair of nodes.}
Let $\mathcal N_i$ denote the neighborhood of agent $i$.  
Agents will be assigned corresponding opinion vectors $\mathbf x^i \in \mathbb R^k$ of which the $j$-th component is denoted $x_j^i$.
We let~$\lVert \cdot \rVert_1$ denote the canonical 1-norm $\lVert \mathbf x \rVert_1 = \sum_{i=1}^{k} |x_i|$. 
The identity matrix of size $k \times k$ is denoted $\mathbb I_k$ with subscript omitted when the size is clear from context.
All inequalities apply element-wise. Vectors of the canonical orthonormal basis in $\mathbb R^k$ are represented by $\mathbf e^i$ where $ i \in \{1, ..., k\}$. When convenient we use the shortened notation $\mathbf{x}^+$, dropping the time argument, to indicate one step of the evolution of some dynamics relative to a previous state~$\mathbf{x}$.

\subsection{Model Setup}

Let $\mathcal G$ be a simple, connected graph with $n$ agents. Each agent $i$ in the graph has a state $\mathbf x^i\in\mathbb{R}^k_{\ge0}$ located on the $(k-1)$-simplex, defined as follows, 
\begin{definition}[k-simplex]
$$\Delta^k = \left \{ \mathbf x \in \mathbb R^{k+1} : \lVert \mathbf x \rVert_1 = 1 , \, \mathbf x \geq \mathbf 0 \right \}.$$
\end{definition}
\vspace{1mm}
The state $\mathbf{x}^i$ represents the \textcolor{black}{relative} opinion of agent~$i$ of the $k$ alternatives. Let $\mathcal{K} = \{1,...,k\}$ denote the set of alternatives corresponding to the indices of $\mathbf{x}^i$. Assign to each agent~$i \in \mathcal V$ a diagonal, nonnegative matrix $R^i \in \mathbb R^{k\times k}_{\ge 0}$. The update rule for every $t \in \mathbb N$ is then given by
% \begin{equation*}
% \mathbf x^i(0) \in \Delta^{k-1}
% \end{equation*}
\begin{equation}\label{eq:update}
    \mathbf x^i (t+1) = \frac{\mathbf x^i(t) + \sum_{j \in \mathcal N_i} R^i \, \mathbf x^j(t)}{\lVert \mathbf x^i(t) + \sum_{j \in \mathcal N_i} R^i \, \mathbf x^j(t) \rVert_1},
\end{equation}
with $\mathbf{x}^i(0) = \mathbf{x}^i_0 \in \Delta^{k-1}$.
The update equation consists of three main components, which we briefly explain here. The first component in the numerator is the opinion of the agent from the last time step. The second term of the numerator captures \textcolor{black}{the filtering of the neighbors' opinions through agent $i$'s bias.} The index corresponding to the largest bias is the most sensitive -- agents have the easiest time moving their opinion in that direction. Note that each agent has an individual bias, in contrast to \cite{Parsegov2017} where one bias between topics is defined for the whole graph.

Finally, the denominator brings the opinion vector back onto $\Delta^{k-1}$, in other words, the opinion vector is normalized. As a result, an increase in one index necessarily causes a decrease in some other index. This is in contrast to standard multidimensional averaging and bounded confidence, where indices evolve independently. \textcolor{black}{This models situations such as the choice between competing products, or probabilistic ones such as the split of an agent's attention between different news sources. The normalization is a mechanism for coupling, and permits more nuanced behaviors.}
%, and can model, for example, a person's relative opinion (e.g. relative trust) of different options, or a split of an agent's attention between different options (e.g. media sources).

% Our choice of domain differs from that of \cite{Parsegov2017}, where opinions take arbitrarily large values. In particular, our choice makes results more interpretable as shares of a limited resource, or probabilities of choosing a certain option. The nonlinear model proposed in \cite{Bizyaeva2023} chooses a different domain that is nonetheless isomorphic to ours. Interpreting the results of their model requires finding a suitable morphism, whilst ours is directly interpretable. A more important difference is a much higher number of parameters and the use of two separate sigmoid nonlinearities in their model, making it harder to analyze.

\begin{remark}
    The bias matrix $R^i$ is uniquely determined by a vector $\mathbf r^i \in \mathbb R^k$ such that $R^i = \text{diag}(\mathbf r^i)$. We will be referring to the vector in places where it aids clarity.
\end{remark}

\begin{remark}\label{remark:obvious}
    Agent $i \in \mathcal V$ can increase its opinion in index $\ell \in \mathcal{K}$ only if the corresponding diagonal entry of its bias matrix $R^i$ is positive in that index and there exists an agent in its neighborhood with a positive opinion in that index.
\end{remark}

% \begin{proof}
%     Look at the update equation for index~$l$ of node~$i$.

%     $$ x^i_l (t+1) = \frac{x^i_l(t) + \sum_{j \in \mathcal N_i} R^i_l \, x^j_l (t) }{\lVert \mathbf x^i(t) + \sum_{j \in \mathcal N_i} R^i \, \mathbf x^j(t)  \rVert_1 } $$

% \noindent
% If either of the conditions are not met, the right-most sum equals zero. In that case, 
%     $$ x^i_l (t+1) = \frac{1}{\lVert \mathbf x^i(t) + \sum_{j \in \mathcal N_i} R^i \, \mathbf x^j(t)  \rVert_1 } \; x^i_l(t).$$

% \noindent
% Since all of the numbers inside the norm are nonnegative and $\mathbf x^i(t) = 1$, we can write  
%     $$ x^i_l (t+1) = \frac{1}{1 + \lVert \sum_{j \in \mathcal N_i} R^i \, \mathbf x^j(t)  \rVert_1 } \; x^i_l(t) \leq x^i_l(t) \, .$$
% \end{proof}

\begin{example}
Let $k=2$. This could be interpreted as a scenario of choosing between two different options. We examine an agent who is currently undecided but prejudiced toward the first option, meaning that
$$ \mathbf x^1 (0) = \begin{bmatrix}
    0.5 \\ 0.5
\end{bmatrix} 
\;\;\; R^1 = \begin{bmatrix}
    0.6 & 0\\
    0 & 0.4 
\end{bmatrix}. $$
We introduce two neighbors with initial opinions ${\mathbf x^2(0)=[0.8 \; 0.2]^\top}$ and ${\mathbf x^3(0)=[0.2 \; 0.8]^\top}$ and calculate the sum in the numerator of the update term for $\mathbf x^1$. 
$$
R^1 \mathbf x^2(0) + R^1 \mathbf x^3(0) = 
\begin{bmatrix}
    0.48\\
    0.08
\end{bmatrix}
+
\begin{bmatrix}
    0.12 \\
    0.32
\end{bmatrix}
$$
After normalization, the updated opinion becomes
$$
\mathbf x^1(1) = 
\begin{bmatrix}
    0.55 \\ 0.45
\end{bmatrix}.
$$
The neighbors were equally sure about their options, but~$\mathbf x^2(0)$ was more in line with the agents's bias; thus it produced a comparatively greater push toward its option.
Further iterations and the steady state of the model depend both on $R^2$, $R^3$ and the structure of the interaction graph, as can be seen from the examples in Section~\ref{section:numerical}.
\end{example}
Next, let us give a simplistic real-world example describing the role of the bias vector $\mathbf{r}$ in the opinion evolution. 
\begin{example}
Consider a case where agents represent individuals whose incomes can be roughly divided in a lower, middle and upper bracket. This might be encoded in their bias vectors with $[1 \; 0 \; 0]$ corresponding to a lower bracket, $[0.5 \: 0.5 \: 0]$ a lower-middle bracket and so on. Agents then form an opinion on which bracket deserves a tax cut in the budget. Their opinion may then align with their bias, but only if presented with that option by their neighbors, that is, if the neighborhood is fairly homogeneous. An upper middle class earner $\mathbf{r}^i = [0 \; 0.6 \; 0.4]$ surrounded by upper class neighbors $\mathbf{r}^j = [0 \; 0 \;  1]$, on the other hand, has a high chance of advocating for the upper class as theirs is the locally dominant opinion. Trust in media with different political leanings may follow a similar logic.  
\end{example}
\color{black}

The following proposition shows that, for certain parameter choices, our model reduces to a simple, well-known averaging model.
\vspace{2pt}

% \begin{example}
%     Assume a fully connected graph. Using the same opinions as before, but setting $R^i = \mathbb I_k \, (\forall i \in \mathcal G)$
%     reduces the model to an averaging model as proved in [TODO]. This means that 
%     $$
%     \mathbf x^i(\infty) = 
%     \begin{bmatrix}
%         0.5 \\ 0.5
%     \end{bmatrix}
%     \; (\forall i \in \mathcal G)
%     $$
%     which is the result that the DeGroot model would predict.
% \end{example}

\begin{proposition}
\label{prop:cons}
    For any $k \in \mathbb N$, $n \in \mathbb N$ and $\mathcal G$ satisfying the definition, setting
    $ R^i = \mathbb I_k \; (\forall i \in \mathcal V) $
    reduces the model to the DeGroot model \cite{degroot1974reaching}.%with uniform weights in each index. 
\end{proposition}
\begin{proof}
Inserting in $R^i = \mathbb I_k$ in (\ref{eq:update}) gives
% \begin{align*}
%     \mathbf x^i (t+1) &= \frac{\mathbf x^i(t) + \sum_{j \in \mathcal N_i} \mathbb I \mathbf x^j(t) }{\lVert \mathbf x^i(t) + \sum_{j \in \mathcal N_i} \mathbb I \, \mathbf x^j(t)  \rVert_1 } \\
%     &= \frac{\mathbf x^i(t) + \sum_{j \in \mathcal N_i} \mathbf x^j(t) }{\lVert \mathbf x^i(t) + \sum_{j \in \mathcal N_i}  \mathbf x^j(t)  \rVert_1 }. \\
% \end{align*}
$$\mathbf x^i (t+1) = \frac{\mathbf x^i(t) + \sum_{j \in \mathcal N_i}  \mathbf x^j(t) }{\lVert \mathbf x^i(t) + \sum_{j \in \mathcal N_i}  \mathbf x^j(t)  \rVert_1 }.$$
Since every $\mathbf x^i$ is nonnegative and unit in the 1-norm, the denominator simplifies, \textcolor{black}{leaving} the averaging formula:
\[\mathbf x^i(t+1) = \frac{\mathbf x^i(t) + \sum_{j \in \mathcal N_i} \mathbf x^j(t) }{1 + | \mathcal N_i |} .\]\end{proof}
\begin{remark}
    In Proposition~\ref{prop:cons}, setting biases to $c \, \mathbb I_k$ for any $c \in \mathbb R_+$ gives the same result, the only difference being the convergence rate. 
\end{remark}

\section{Analysis} \label{anaylsis}

% \begin{lemma}
%     Let $$ f : \Delta^k \rightarrow \Delta^k $$
%     $$ f(\mathbf x(t)) = \mathbf x(t+1) $$
%     $f$ is continuous.
% \end{lemma}
% \begin{proof}

% Write out the update equation.
% $$ \mathbf x^i (t+1) = \frac{\mathbf x^i(t) + \sum_{j \in \mathcal N_i} R^i \, \mathbf x^j(t) }{\lVert \mathbf x^i(t) + \sum_{j \in \mathcal N_i} R^i \, \mathbf x^j(t)  \rVert_1 }  $$
% The update rule consists of continuous operations, possibly with the exception of division if the denominator of the fraction turns out to be zero. Suppose that it is.
% $$ \lVert \mathbf x^i(t) + \sum_{j \in \mathcal N_i} R^i \, \mathbf x^j(t)  \rVert_1 = 0 $$
% As all of the summands are nonnegative, they have to be equal to zero. This now gives
% $$ \lVert \mathbf x^i(t) \rVert_1 = 0 $$
% which is in contradiction with 
% $$ \lVert \mathbf x^i(t) \rVert_1 = 1.$$
% Therefore, the denominator cannot be zero and $f$ is continuous.
% \end{proof}

In this section, we present \textcolor{black}{certain} analytical results pertaining to the fixed points of the system. While these are difficult to characterize completely due to the possible diversity of outcomes depending on the network topology, bias vectors and initial opinions, we  give conditions for a number of \textcolor{black}{relevant} special cases. \textcolor{black}{ The objective is to characterize the mathematical properties of the model itself, to inform its applicability in practical contexts. The following is a technical result on trivial fixed points. }
\begin{proposition}\label{prop:simplefixed}
    Configurations where all agents share the same state with a 1 in one entry and a 0 in all other entries
    $$\mathbf x^1 , \cdots , \mathbf x^n = 
    \begin{bmatrix}
        0 ,
        ... ,
        0 ,
        1 ,
        0 ,
        ... ,
        0
    \end{bmatrix}^\top
    $$
    \noindent
    are fixed points of (\ref{eq:update}).
\end{proposition}

\begin{proof}
    Suppose $\mathbf x^i = \mathbf e^\ell$ for all $i \in \mathcal V$ and some $l \in \mathcal{K}$. The result follows from insertion in (\ref{eq:update}).
    %Then, for any $i \in \mathcal V$:
%\begin{align*}
% \mathbf x^i (t+1) %&= \frac{\mathbf x^i(t) + \sum_{j \in \mathcal N_i} R^i \, \mathbf x^j(t)}{\lVert \mathbf x^i(t) + \sum_{j \in \mathcal N_i} R^i \, \mathbf x^j(t)  \rVert_1 }   \\ 
 %&= \frac{\mathbf e^\ell + \sum_{j \in \mathcal N_i} R^i \, \mathbf e^\ell}{\lVert \mathbf e^\ell + \sum_{j \in \mathcal N_i} R^i \, \mathbf e^\ell  \rVert_1 }  \\
 %&= \frac{\left [ 0, ..., 0, 1 + |\mathcal N_i| \, r^i_{\ell}, 0, ..., 0 \right ]^\top}{\lVert \left [ 0, ..., 0, 1 + |\mathcal N_i| \, r^i_{\ell}, 0, ..., 0 \right ]^\top \rVert_1 } 
% = \mathbf x^i(t). 
%\end{align*}
\end{proof}
%\textcolor{black}{While not necessarily sociologically relevant, the result serves as a sanity check.}

\begin{example}
    These fixed points can easily be shown to be attractive for certain trivial configurations of the model. Let $k \in \mathbb N$ and $n \in \mathbb N$. Then, if there exists an index $\ell$ such that $R^i=\text{diag}(\mathbf e^\ell)$ for all agents $i \in \mathcal{V}$, it holds that $\mathbf x^i (t) \rightarrow \mathbf e^\ell$ for all $i \in \mathcal{V}$ and for almost all initial conditions. 
\end{example}

\subsection{General fixed points and bias with zero elements}

\edit{We next give a general result characterizing }%The following proposition is a general result that characterizes 
the fixed points of \eqref{eq:update}. In cases where zero bias toward some options is allowed, agents may stop communicating with neighbors that have certain opinions. This in turn allows for fixed points where the opinions of some agents are not uniquely determined by neighbors' opinions.
%\emma{I still have some trouble with the preceding sentence and what it means. It sounds a bit chaotic and not desirable for a model...}

\begin{proposition}\label{prop:localize}
    At the fixed points of the dynamics \eqref{eq:update}, the state $\mathbf{x}^i$ corresponding to an agent $i \in \mathcal V$ satisfies exactly one of the following two conditions:
    % \begin{equation*}
    %     (i)\; ||\sum\limits_{j\in\mathcal{N}_i}R^i\mathbf{x}^j||_1 = 0,\;\; (ii)\; \mathbf{x}^i = \frac{1}{||\sum\limits_{j\in\mathcal{N}_i}R^i\mathbf{x}^j||_1}\sum\limits_{j\in\mathcal{N}_i}R^i\mathbf{x}^j
    % \end{equation*}
    \vspace{1mm}
    \begin{itemize}
        \item[$(i)$] $||\sum_{j\in\mathcal{N}_i}R^i\mathbf{x}^j||_1 = 0$
        \vspace{1mm}
        \item[$(ii)$] $\mathbf{x}^i = \frac{1}{||\sum_{j\in\mathcal{N}_i}R^i\mathbf{x}^j||_1}\sum\limits_{j\in\mathcal{N}_i}R^i\mathbf{x}^j$
    \end{itemize}
    \vspace{1mm}
\end{proposition}

\begin{proof}
For all fixed points of \eqref{eq:update} it must hold that $(\mathbf{x}^i)^+ = \mathbf{x}^i$, which is equivalent to
\begin{align*}
    \frac{\mathbf{x}^i(1-||\mathbf{x}^i + \sum\limits_{j\in\mathcal{N}_i}R^i\mathbf{x}^j||_1) + \sum\limits_{j\in\mathcal{N}_i}R^i\mathbf{x}^j}{||\mathbf{x}^i + \sum\limits_{j\in\mathcal{N}_i}R^i\mathbf{x}^j||_1} = 0\\
    \iff \mathbf{x}^i||\sum\limits_{j\in\mathcal{N}_i}R^i\mathbf{x}^j||_1 = \sum\limits_{j\in\mathcal{N}_i}R^i\mathbf{x}^j,
\end{align*}
\noindent
where division by $||\sum_{j\in\mathcal{N}_i}R^i\mathbf{x}^j||_1$ yields the expression ($i$) for~$\mathbf{x}^i$ at the fixed point unless $||\sum_{j\in\mathcal{N}_i}R^i\mathbf{x}^j|| = 0$.
\end{proof}

\begin{corollary}\label{cr:zero}
    Let the bias $\textbf{r}^i$ of some agent $i$ have a zero element. Then, at the fixed points of~\eqref{eq:update}, either $\mathbf{x}^i_{\ell} = 0$ or~($i$) in Proposition~\ref{prop:localize} holds for this agent.
\end{corollary}
\vspace{0.5mm}

Corollary~\ref{cr:zero} is a direct consequence of Proposition~\ref{prop:localize} in the case where agents have no susceptibility to some option $\ell$. It means that in any fixed configuration the agents either reject the option they are not susceptible to ($\mathbf{x}_{\ell_i} = 0$), or they are no longer influenced by their neighbors. 

\begin{corollary}\label{cr:boundary}
    Let $\textbf{r}^i$ have strictly positive entries for all agents $i$. At the fixed points of~\eqref{eq:update}, if any agent is on the boundary of $\Delta^{k-1}$ (that is, $\mathbf{x}^i_{\ell} = 0$ for at least one index~${\ell\in\mathcal{K}}$) then all other agents must be at the same face.
\end{corollary}

\begin{proof}
    Condition ($i$) in Proposition~\ref{prop:localize} can never hold under this assumption on the biases $\textbf{r}^i$. Evaluating the expression in point~($ii$) of Proposition \ref{prop:localize} we see that $\mathbf{x}^i_\ell$ can only be zero at a fixed point if all neighbors $j\in\mathcal{N}_i$ have~${\mathbf{x}^j_\ell = 0}$. Repeating the argument for these neighbors yields that $x^i_\ell = 0$ must be true for all agents.
\end{proof}

\subsection{Rejection of non-dominant alternatives}
%\subsection{Rejection of unpopular alternatives}
In order to present the next result, we first introduce a partition of the set of alternatives $\mathcal{K}$ into dominant and recessive opinions. 
\begin{definition}\label{def:dominant}
    Let $\mathcal{K}$ be partitioned into two subsets; the dominant set $\mathcal{D}$ and the recessive set $\mathcal{L}$. The recessive set is defined as the largest set $\mathcal{L}\subsetneq\mathcal{K}$ such that for all $\ell\in\mathcal{L}$ and~${d\in\mathcal{K}\setminus\mathcal{L}}$ the inequality $r^i_\ell < r^i_d$ holds for all agents $i$. The dominant set is defined as the remainder, $\mathcal{D} = \mathcal{K}\setminus\mathcal{L}$.
\end{definition}
% \emma{I think we should add: ``Note that $\mathcal{D}$ might be the empty set''}
\begin{remark}
    Note that the largest set $\mathcal{L}$ is uniquely defined (though it may be empty) and all smaller sets fulfilling the criteria are subsets of $\mathcal{L}$.
\end{remark}

We can now show that opinions in the set $\mathcal{L}$ are suppressed by the dynamics.
% \emma{Not true if $\mathcal{L} = \mathcal{K}$?!}
\begin{proposition}\label{prop:dominant}
   At the fixed points of \eqref{eq:update}, for almost all initial values, all agents $i$ have opinion $\mathbf{x}^i_\ell = 0$ for $\ell\in\mathcal{L}$. 
\end{proposition}

\begin{proof}
    We define the Lyapunov candidate 
    \begin{equation}
        V(\mathbf{x}) = \max_i \sum_{\ell\in\mathcal{L}} x^i_\ell 
    \end{equation} 
    which satisfies $V(\mathbf{x}) = 0$ only when, for every agent $i$, ${x^i_\ell = 0}$ for all $\ell\in\mathcal{L}$. We wish to show that
    \begin{equation}\label{eq:lyap}
        V(\mathbf{x}^+)-V(\mathbf{x}) < 0
    \end{equation}
    on the domain 
    \begin{equation}\label{eq:domain}
        \{\mathbf{x} : \mathbf{x}^i\in\Delta^{k-1}, V(\mathbf{x}) \ne 1\}
    \end{equation}
    which is invariant under the dynamics \eqref{eq:update}. Let ${\Bar{x}_\mathcal{L} := \max\limits_{i} (\sum_{\ell\in\mathcal{L}}x^i_{\ell})}$. Inserting \eqref{eq:update} into \eqref{eq:lyap} gives
    \begin{align*}
        V(\mathbf{x}^+)\!-\!V(\mathbf{x}) &= \max_i\sum_{\ell\in\mathcal{L}}\left(\frac{x^i_{\ell}+\sum\limits_{j\in\mathcal{N}_i}r^i_\ell x^j_\ell}{1+||\sum\limits_{j\in\mathcal{N}_i}R^i\mathbf{x}^j||_1}\right)-\Bar{x}_\mathcal{L}\\
        &\le \max_i\frac{\Bar{x}_\mathcal{L} + \sum\limits_{\ell\in\mathcal{L}}\sum\limits_{j\in\mathcal{N}_i}r^i_\ell x^j_\ell}{1+||\sum\limits_{j\in\mathcal{N}_i}R^i\mathbf{x}^j||_1}-\Bar{x}_\mathcal{L}\\
        &= \max_i\frac{\sum\limits_{\ell\in\mathcal{L}}\sum\limits_{j\in\mathcal{N}_i}r^i_\ell x^j_\ell-\Bar{x}_\mathcal{L}||\sum\limits_{j\in\mathcal{N}_i}R^i\mathbf{x}^j||_1}{1 + ||\sum\limits_{j\in\mathcal{N}_i}R^i\mathbf{x}^j||_1}.
    \end{align*}
    If we can show that the expression being maximized is strictly negative for all $i$ then \eqref{eq:lyap} must also hold. This yields the condition
    %\begin{align*}
    ${\sum_{\ell\in\mathcal{L}}\sum_{j\in\mathcal{N}_i}r^i_\ell x^j_\ell < \Bar{x}_\mathcal{L}\sum_{j\in\mathcal{N}_i}||R^i\mathbf{x}^j||_1}.$
    %\end{align*}
    A more strict condition, implying the above, is given by the element-wise expression
    %\begin{align*}
    ${\sum_{\ell\in\mathcal{L}}r^i_\ell x^j_\ell < \Bar{x}_\mathcal{L}||R^i\mathbf{x}^j||_1}$.
    %\end{align*}
    Let $\rho^i = \min_{\ell\in\mathcal{D}}r^i_\ell$. Division by $\rho^i$ yields
    \begin{align}
        \sum_{\ell\in\mathcal{L}} \frac{r^i_\ell}{\rho^i} x^j_\ell &< \frac{1}{\rho^i}\Bar{x}_\mathcal{L}||R^i\mathbf{x}^j||_1\nonumber\\
        %\iff \sum_{\ell\in\mathcal{L}} \frac{r^i_\ell}{\rho^i} x^j_\ell &< \Bar{x}_\mathcal{L}\left(\sum\limits_{\ell\in\mathcal{L}} \frac{r^i_\ell}{\rho^i} x^j_\ell + \sum\limits_{\ell\in\mathcal{D}} \frac{r^i_\ell}{\rho^i} x^j_\ell \right)\nonumber\\
        \iff(1-\Bar{x}_\mathcal{L})\sum_{\ell\in\mathcal{L}} \frac{r^i_\ell}{\rho^i} x^j_\ell &< \Bar{x}_\mathcal{L}\sum\limits_{\ell\in\mathcal{D}} \frac{r^i_\ell}{\rho^i} x^j_\ell.\label{eq:ineq}
    \end{align}
    Note that according to Definition \ref{def:dominant} we have $r^i_\ell/\rho^i < 1$ for~${\ell\in\mathcal{L}}$ and $r^i_\ell/\rho^i \ge 1$ for~$\ell\in\mathcal{D}$. As a consequence of the definition of $\Bar{x}_\mathcal{L}$, the factors of \eqref{eq:ineq} satisfy %the individual inequalities
    $$
        (1-\Bar{x}_\mathcal{L}) \le \sum\limits_{\ell\in\mathcal{D}} \frac{r^i_\ell}{\rho^i} x^j_\ell
    \quad\text{and}\quad 
        \sum_{\ell\in\mathcal{L}} \frac{r^i_\ell}{\rho^i} x^j_\ell < \Bar{x}_\mathcal{L},
    $$
    for all $\mathbf{x}^i\in\Delta^{k-1}$ such that $\sum_{\ell\in\mathcal{L}} x^i_\ell\ne1$, thus implying~\eqref{eq:ineq}. This proves that $V$ is strictly decreasing on the domain stated in \eqref{eq:domain}, concluding the proof.
    \vspace{1mm}
\end{proof}

Proposition \ref{prop:dominant} is a versatile result, giving formal motivation for several intuitive properties of the model. The most obvious is the case of aligned biases, where $\mathbf{r}^i = \mathbf{r}$ (or when $\mathbf{r}$ has the same ordering of alternatives). In this case the largest proper subset $\mathcal{L}$ in Definition \ref{def:dominant} contains all alternatives except the one with highest bias. Application of Proposition \ref{prop:dominant} to this instance yields a consensus for the alternative favored by the biases, which is the natural conclusion. Further, in cases where all agents are biased against a certain set of alternatives $\mathcal{L}$, these are suppressed by the dynamics and not present in the final opinion distribution. This happens despite the fact that the agents may not agree on the best option, and regardless of whether some agents initially held a favorable opinion of the alternatives in $\mathcal{L}$.

%\emma{I might suggest adding a corollary based on this text. ie consensus is reached at $x_i_j = 1$ if the option is dominating for all. We can then maybe say that even if a single agents has a different ordering, there will be no consensus (I'm sure that can be shown too?). Let's ask David to write that? I think it is important to show that we actually have quite a few results.}

\begin{figure*}[ht!]
    % \centering
    % \vspace{-30pt}
    % \includegraphics[width=0.6\linewidth]{polarization/x0.pdf}
    \vspace{-10pt}
    % \caption{Initial distribution of opinions}
    % \label{fig:polx0}

    \centering
    \begin{minipage}{0.32\textwidth}
        \centering
        \includegraphics[width=.9\textwidth]{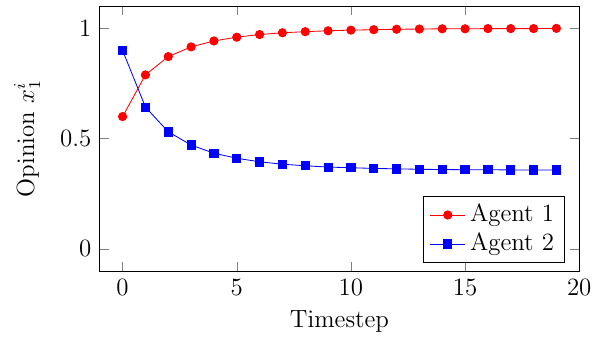} % second figure itself
        \subcaption{the maximal bias case}
        \label{fig:con}
    \end{minipage}\hfill
    \begin{minipage}{0.32\textwidth}
        \centering
        \includegraphics[width=.9\textwidth]{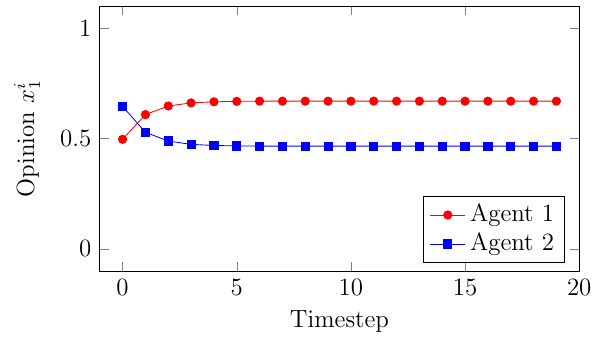} % first figure itself
        \subcaption{mediation}
        \label{fig:med}
    \end{minipage}\hfill
    \begin{minipage}{0.32\textwidth}
        \centering
        \includegraphics[width=.9\textwidth]{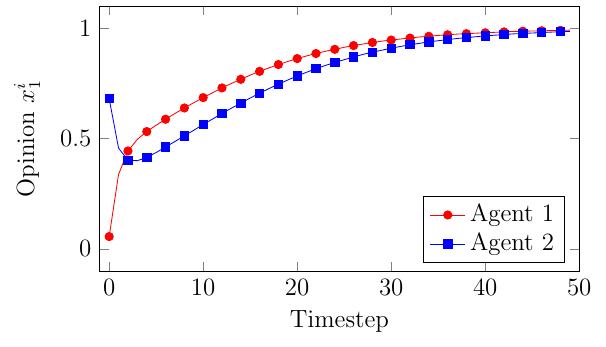} % second figure itself
        \subcaption{agent 2 influenced toward option 1}
        \label{fig:rad}
    \end{minipage}\hfill
    \caption{The $n=k =2$ case, despite its simplicity, showcases different fixed points depending on the biases and starting opinions of the agents. Biasing agents completely toward opposing options leads to a race to the boundary of the domain, as seen in (a). The symmetric but non-maximal biases in (b) give fixed points in the interior of the simplex. Finally, (c) showcases that agents with neutral biases can be influenced toward either option.}
    \label{fig:ex1}
\end{figure*}

\subsection{Fixed point stability analysis for $n=2$, $k=2$}
\label{2agentfixed}
%This subsection examines the simplest non-trivial case with~$n=2$ agents and  opinion dimension~$k=2$. 
\edit{Next, we examine in further detail the simplest non-trivial case of~$n=2$ agents and opinion dimension~$k=2$. }
\textcolor{black}{Due to a reasonable number of parameters, this scenario lends itself to direct analysis. }% which can then be carried over to larger dimensions. }
\edit{Now, }since $x^1_1 + x^1_2 = 1$, examining one component per agent is sufficient and we let $x^i$ denote $x^i_1$. Let ${\mathbf r^1 = [ \alpha^1 \; \beta^1]^\top}$ and ${\mathbf r^2 = [\alpha^2 \; \beta^2]^\top}$. The dynamics~\eqref{eq:update}~become:
% \begin{equation} \label{eq:x1}
% x^1 (t+1) = \frac{x^1(t) + \alpha^1 x^2(t)}{1 + \beta^1 + (\alpha^1 - \beta^1)x^2(t) }
% \end{equation}
% \begin{equation} \label{eq:x2}
% x^2 (t+1) = \frac{x^2(t) + \alpha^2 x^1(t)}{1 + \beta^2 + (\alpha^2 - \beta^2)x^1(t) }
% \end{equation}
\begin{subequations}
    \begin{align}
x^1 (t+1) &= \frac{x^1(t) + \alpha^1 x^2(t)}{1 + \beta^1 + (\alpha^1 - \beta^1)x^2(t) } \label{eq:x1} \\
x^2 (t+1) &= \frac{x^2(t) + \alpha^2 x^1(t)}{1 + \beta^2 + (\alpha^2 - \beta^2)x^1(t) } \label{eq:x2}
    \end{align}
\end{subequations}
of which the fixed points are given by:
%\[x^{1*}\! =\! \frac{\alpha^1 x^{2*}}{\alpha^1 x^{2*} + \beta^1(1-x^{2*})},~x^{2*} \! =\! \frac{\alpha^2 x^{1*}}{\alpha^2 x^{1*} + \beta^2(1-x^{1*})}.\]
\begin{subequations}
    \begin{align}
 x^{1*} &= \frac{\alpha^1 x^{2*}}{\alpha^1 x^{2*} + \beta^1(1-x^{2*})} \label{eq:x1fix} \\
 x^{2*} &= \frac{\alpha^2 x^{1*}}{\alpha^2 x^{1*} + \beta^2(1-x^{1*})}.
    \end{align} \label{eq:x2fix}
\end{subequations}
 % \begin{equation} \label{eq:x1fix}
 % x^{1*} = \frac{\alpha^1 x^{2*}}{\alpha^1 x^{2*} + \beta^1(1-x^{2*})}
 % \end{equation}
 % \begin{equation} \label{eq:x2fix}
 % x^{2*} = \frac{\alpha^2 x^{1*}}{\alpha^2 x^{1*} + \beta^2(1-x^{1*})}.
 % \end{equation}
Substituting (\ref{eq:x2fix}) into (\ref{eq:x1fix}) gives a quadratic equation for the fixed points of agent 1:
\begin{equation} \label{eq:fixie}
\left(\alpha^1 \alpha^2 - \beta^1 \beta^2\right)\left((x^{1*})^2 - x^{1*})\right) = 0.
\end{equation}
To determine local asymptotic stability of the fixed points, we will make use of the following lemma.
\begin{lemma}\label{lemma:schur}
    A nonnegative matrix $\left [\begin{smallmatrix}
        a & b \\
        c & d
    \end{smallmatrix} \right ]
    $ is Schur stable if and only if $a+d<2$ and $a+d+bc < 1+ ed$.
\end{lemma}
\begin{proof}
    By the Perron-Frobenius theorem, the dominant eigenvalue is real and positive. Explicit computation gives $\lambda_{\text{max}} = \frac{1}{2}\left( a+d+ \sqrt{(a+d)^2 -4ad + 4bc}\right)$. Finally, $\lambda_{\text{max}} < 1$ is equivalent to the statement in the lemma.
\end{proof}

We can now characterize the fixed points for this low-dimensional case: % By combining the results attained so far, the following proposition characterizes some of the fixed points observed in low dimensional instances.
\begin{proposition}\label{prop:giacomo}
The fixed points of (\ref{eq:x1})--(\ref{eq:x2}) depend on the bias  vectors  ${\mathbf r^1 \! = [ \alpha^1 \; \beta^1]^\top}$ and ${\mathbf r^2 = [\alpha^2 \; \beta^2]^\top}$  as~follows:
\begin{itemize}
    \item[$(i)$] if $\alpha^1 \alpha^2 < \beta^1 \beta^2$ then (0,0) is a locally asymptotically stable fixed point while (1,1) is an unstable fixed point, 
    \item[$(ii)$] if $\alpha^1 \alpha^2 > \beta^1 \beta^2 $ then (1,1) is a locally asymptotically stable fixed point while (0,0) is an unstable fixed point, 
    \item[$(iii)$] if $\alpha^1 \alpha^2 = \beta^1 \beta^2 $ then there exists a continuum of fixed points within $\Delta^1$.
\end{itemize}
\end{proposition}

% \begin{proposition}\label{prop:giacomo}
% The fixed points of the system given by (\ref{eq:x1}) and (\ref{eq:x2}) depend on the biases as follows:
% \begin{itemize}
%     \item[(i)] $\alpha^1 \alpha^2 < \beta^1 \beta^2 \implies (0,0) \text{ LAS, } (1,1) \text{ unstable}$
%     \item[(ii)] $\alpha^1 \alpha^2 > \beta^1 \beta^2 \implies (1,1) \text{ LAS, } (0,0) \text{ unstable}$
%     \item[(iii)] $\alpha^1 \alpha^2 = \beta^1 \beta^2 \implies \text{a continuum of fixed points}$
% \end{itemize}
% \end{proposition}

\begin{proof}
The fixed points in cases $(i)$ and $(ii)$ are easily obtained from~\eqref{eq:fixie}.
%It is evident from equation (\ref{eq:fixie}) that these are the fixed points in case (i) and (ii). 
%\emma{I agree for (1,1) and (0,0), but the continuum case (iii) is not really evident...Plus you probably need to remind the reader that $x^1,x^2$ live in the domain [0,1]}. 
To determine stability, examine the Jacobian of the system at~(0,0):
$$ \mathbf J(0,0) = \begin{bmatrix}
    \frac{1}{1+\beta^1} & \frac{\alpha^1}{1 + \beta^1} \\
    \frac{\alpha^2}{1+ \beta^2} & \frac{1}{1 + \beta^2 }
\end{bmatrix}.$$
Applying Lemma \ref{lemma:schur} gives that the fixed point $(0,0)$ is locally asymptotically stable if and only if $\alpha^1 \alpha^2 < \beta^1 \beta^2$. By symmetry, $(1,1)$ is stable if and only if $\alpha^1 \alpha^2 > \beta^1 \beta^2$. In case $(iii)$ \eqref{eq:fixie} reduces to $0=0$ implying no additional criteria for the fixed points other than (\ref{eq:x1fix}) and~ (\ref{eq:x2fix}).
\end{proof}

In short, agents converge toward their combined preference unless they, as a pair, happen to be equally predisposed toward both options. 
%Numerical evidence suggests convergence is almost global in this simple scenario.
\textcolor{black}{Numerical evidence suggests that larger networks exhibit similar phenomena, where the choice of biases affects the cardinality of the set of fixed points.}

%\vspace{-5pt}
\section{Numerical examples}
\label{section:numerical}
This section presents two numerical examples; the first to provide intuition of the model's behaviors through the simple two-agent two-dimensional case, the second to illustrate the interesting phenomenon of polarization versus bias mediation in a larger network. 

\subsection{Two Agents}
\label{ex:two}

%\textcolor{black}{To complement the analysis done in Section~\ref{2agentfixed}, we use numerical experiment with the goal of fully describing the smallest relevant case.}
\edit{In order to illustrate the behaviors described in Section~\ref{2agentfixed} for the smallest non-trivial case, we provide three examples with qualitatively different behaviors.  }

\subsubsection*{1) $\alpha^1 \alpha^2 = \beta^1 \beta^2$}
In line with Proposition \ref{prop:giacomo}, these combinations of biases admit a continuum of fixed points. First, in Fig.~\ref{fig:con}, an instance with $\mathbf r^1 = [1 \; 0]^\top$ and $\mathbf r^2 = [0 \; 1]^\top$ is shown. The agents' preconceptions are completely incompatible, and their opinions move toward their biases and away from each other. This also serves as an illustration of Corollary~\ref{cr:zero}, where zeros in bias vectors result in fixed points on the boundary \edit{or agents ceasing to influence each other. }

Fig.~\ref{fig:med} showcases similar dynamics, the difference being that a choice of non-maximal yet symmetrical biases ${\mathbf r^1 = [0.7 \; 0.3]^\top}$ and ${\mathbf r^2 = [0.3 \; 0.7]^\top}$ results in a fixed point in the interior of the simplex. The two agents mediate, each moving their opinion vector toward their bias vector but stopping before reaching the boundary.

% \subsubsection*{a) Maximally Biased Agents}
%Agent 1 is assigned a bias of $\mathbf r^1 = [1, 0]$ whilst agent 2 is assigned $\mathbf r^2 = [0, 1]$. Their preconceptions being completely incompatible, their opinions move toward their biases and away from each other. 

%In line with Remark \ref{remark:obvious}, the dynamics evolve until one of the agents reaches an edge of the simplex, at which moment the fixed point is reached with the other agent being in-between its starting point and the point on the edge of the simplex corresponding to its bias vector. The result of one such instance is presented in Fig.~\ref{fig:con}. The exact value of this fixed point is not described by any of the theorems, but its existence is consistent with Proposition \ref{prop:simplefixed}.

% The choice of initial opinion impacts the steady state. The longer it takes for the first agent to reach the domain's boundary, the closer the second agent gets to its bias. 
% We can numerically show that if the initial opinion is identical and equally far from the bias vector, i.e. $\mathbf x^1(0) = \mathbf x^2(0) = [0.5, 0.5]$, then both agents asymptotically align with their biases.

% \subsubsection*{b) Mediation}
% Relaxing the previous case, agents can be assigned opposite but non-maximal biases. Fig.~\ref{fig:med} shows such a case with $\mathbf r^1 = [0.7, 0.3]$ and $\mathbf r^2 = [0.3, 0.7]$. The two agents mediate, each moving their opinion vector toward their bias but stopping at some point in the interior of~$\Delta^1$. 

\subsubsection*{2) $\alpha^1 \alpha^2 \neq \beta^1 \beta^2$}
Moving the bias of agent 2 closer to that of agent 1 further changes the behavior of the model. Fig. \ref{fig:rad} shows a simulation with ${\mathbf r^1 = [0.7 \; 0.3]^\top}$ and ${\mathbf r^2 = [0.45 \; 0.55]^\top}$. Having a relatively neutral bias, agent 2 is easily influenced by its neighbor. Applying Proposition~\ref{prop:giacomo}, we see that, since ${\alpha^1 \alpha^2 = 0.7 \cdot 0.45 < 0.3 \cdot 0.55 = \beta^1 \beta^2}$, $(1,1)$ is the locally asymptotically stable fixed point. This \textcolor{black}{agrees with} the simulation, as both agents converge to option~1.

% \begin{figure}[!h]
%     \centering
%     \includegraphics[width=0.9\linewidth]{2agents/rad2.pdf}
%     \caption{Node 2 getting influenced toward option 2}
%     \label{fig:rad2}
% \end{figure}
\begin{figure*}[ht!]
    % \centering
    \vspace{-30pt}
    % \includegraphics[width=0.6\linewidth]{polarization/x0.pdf}
    % \vspace{-25pt}
    % \caption{Initial distribution of opinions}
    % \label{fig:polx0}

    \centering
    \begin{minipage}{0.32\textwidth}
        \centering
        \includegraphics[width=0.7\textwidth]{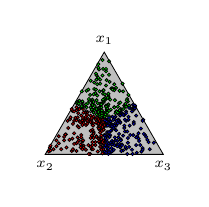} % first figure itself
        \vspace{-20pt}
        \subcaption{starting distribution}
        \label{fig:spstart}
    \end{minipage}\hfill
    \begin{minipage}{0.32\textwidth}
        \centering
        \includegraphics[width=0.7\textwidth]{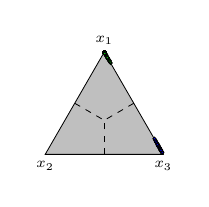} % second figure itself
        \vspace{-20pt}
        \subcaption{spatially correlated biases}
        \label{fig:sppol}
    \end{minipage}\hfill
    \begin{minipage}{0.32\textwidth}
        \centering
        \includegraphics[width=0.7\textwidth]{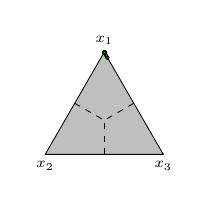} % second figure itself
        \vspace{-20pt}
        \subcaption{uniformly distributed biases}
        \label{fig:spint}
    \end{minipage}
    \caption{Networks with communities such as Watts-Strogatz \textcolor{black}{networks can showcase interesting behaviors. }%lead to complex effects.
    A uniform starting distribution \textcolor{black}{of opinions  as in} (a) can result in a polarized steady state (b) when biases are spatially correlated, that is, if certain small communities are assigned a bias that is opposite to the majority. On the other hand, assigning the same number of {contrarian} biases to individuals chosen uniformly at random across the network leads to agreement, as can be seen in~(c). The figures display the results of a simulation done on a $n=500$ agent network. 
    }
    \label{fig:whole}
    \vspace{-15pt}
\end{figure*}

\subsection{Spatial Correlation and Polarization}

\edit{For a larger-scale example, }consider a connected Watts-Strogatz network~\cite{Watts1998} as the underlying graph. A characteristic of such graphs is the formation of communities -- subsets of agents where intragroup connections are stronger than those with the rest of the graph. Supposing they are given similar biases, such communities can turn into echo chambers under our model, effectively reinforcing each others' biases and ignoring the majority opinion.

To run a numerical experiment, a Watts-Strogatz graph of $n=500$ agents was generated using NetworkX %\cite{networkx} 
and a community detection algorithm 
%\cite{Clauset2004-ae} 
was applied. This revealed three communities of size 226, 222, and 52 within the particular graph. Initial opinions of dimension $k = 3$ were uniformly drawn from the simplex, as can be seen in Fig.~\ref{fig:spstart}. Two scenarios were then analyzed.

\subsubsection*{1) Spatially correlated biases}
All agents in the two larger communities are given a bias of $\mathbf r_A = [0.8 \; 0.09 \; 0.11]^\top$. This is a way of forming a "majority configuration" in the graph. Agents in the smallest community are given a bias of $\mathbf r_B =  [0.11 \; 0.09 \; 0.8]^\top.$

As can be seen from the biases, communities agree in having least preference for option 2, but disagree in which option is most favorable. The steady state, as can be seen in Fig.~\ref{fig:sppol}, is a polarization of opinions. The majority groups go toward option 1, and the minority goes toward option 2. Note that the steady state is not a simple split consensus, as final opinions are actually spread around the corner points of the simplex.

\subsubsection*{2) Randomly assigned biases}
Once again, a majority and a minority group are chosen, this time completely at random. The group sizes match with those in the previous scenario, group A with 448 agents and  group B with 52 agents. As in the previous scenario, groups are assigned biases $\mathbf r_A$ and $\mathbf r_B$, respectively.
The steady state can be seen in Fig.~\ref{fig:spint}. 
Agents in the steady state agree in the sense of choosing an option, but stay spread around the corner of the simplex instead of reaching consensus. Individual biases  still play a role in the final opinion, albeit a comparatively smaller one than in the previous scenario. 
The interpretation here is that the majority manages to integrate the minority, as they are no longer in an echo chamber but are well connected to the rest of the network. \textcolor{black}{While these behaviors are not surprising in light of the model's mechanisms, our model offers a framework for deriving and describing these known behaviors. } %they offer a mechanistic explanation and model for known behaviors.}

\section{Conclusions}
\label{sec:conclusion}
We have proposed an opinion-dynamical model that describes individuals' relative opinion of multiple coupled alternatives, depending on an individual bias, and connections over a social network. Despite the apparent simplicity of the model~\eqref{eq:update}, it is expressive, as showcased through our numerical examples. In particular, simulation on the Watts-Strogatz network manages to capture the polarization and integration phenomena appearing in the social science literature that motivated the model.  We analytically show sensible properties in choice-making scenarios, such as the rejection of alternatives if they are unfavorable to all agents. 

There exist several directions for future work. First, the model can be extended in various directions; e.g. to consider a full bias matrix, allowing further coupling of different options, or a weighted graph to allow for finer specification of relationships between agents. Another direction is stochasticity in opinion exchanges, in order to adjust for imperfect communication often occurring in real discussions. \textcolor{black}{Taking the initial analysis on fixed points in this work as a starting point, we are also  interested in further classification of the set of fixed points and their stability in a general case, along with concrete results on covergence of the model.} Finally, the validation of the model on empirical data on trust in government and media is ongoing.

% The numerical examples showcase the expressivity of our model regardless of network size. The simulation on the Watts-Strogatz network specifically manages to capture the polarization and integration phenomena appearing in the literature which motivated the model. We analytically show sensible properties in choice-making scenarios, such as the rejection of alternatives if they are unfavorable to all agents. There exist several directions for future work. One is to consider a full bias matrix, allowing further coupling of different options. Another is making opinion exchanges stochastic, in order to adjust for imperfect communication often occurring in real discussions. Finally, a weighed graph could be considered to allow for finer specification of relationships between agents.

\section*{Acknowledgement}
We %would like to 
thank Mark Jeeninga, Jonas Hansson and Richard Pates for %a number of 
fruitful discussions related to 
this work. 

\bibliographystyle{IEEEtran}
\bibliography{references}

\begin{thebibliography}{10}
\providecommand{\url}[1]{#1}
\csname url@rmstyle\endcsname
\providecommand{\newblock}{\relax}
\providecommand{\bibinfo}[2]{#2}
\providecommand\BIBentrySTDinterwordspacing{\spaceskip=0pt\relax}
\providecommand\BIBentryALTinterwordstretchfactor{4}
\providecommand\BIBentryALTinterwordspacing{\spaceskip=\fontdimen2\font plus
\BIBentryALTinterwordstretchfactor\fontdimen3\font minus \fontdimen4\font\relax}
\providecommand\BIBforeignlanguage[2]{{%
\expandafter\ifx\csname l@#1\endcsname\relax
\typeout{** WARNING: IEEEtran.bst: No hyphenation pattern has been}%
\typeout{** loaded for the language `#1'. Using the pattern for}%
\typeout{** the default language instead.}%
\else
\language=\csname l@#1\endcsname
\fi
#2}}

\bibitem{eu2019barameter}
\BIBentryALTinterwordspacing
Eurobarometer, ``Standard {E}urobarometer 92, media use in the {E}uropean {U}nion,'' 2019. [Online]. Available: \url{https://europa.eu/eurobarometer/surveys/detail/2255}
\BIBentrySTDinterwordspacing

\bibitem{Schranz2018}
M.~Schranz, J.~Schneider, and M.~Eisenegger, \emph{Media Trust and Media Use}.\hskip 1em plus 0.5em minus 0.4em\relax Springer Fachmedien Wiesbaden, 2018, p. 73–91.

\bibitem{Nickerson1998}
R.~S. Nickerson, ``Confirmation bias: A ubiquitous phenomenon in many guises,'' \emph{Rev Gen Psychol}, vol.~2, no.~2, p. 175–220, June 1998.

\bibitem{Mullainathan2005}
S.~Mullainathan and A.~Shleifer, ``The market for news,'' \emph{Am Econ Rev}, vol.~95, no.~4, pp. 1031--1053, 2005.

\bibitem{Urman2019}
A.~Urman, ``Context matters: political polarization on {Twitter} from a comparative perspective,'' \emph{Media Cult and Soc}, vol.~42, no.~6, p. 857–879, Oct. 2019.

\bibitem{GonzlezBailn2023}
{\relax González-Bailón}.~et~al., ``Asymmetric ideological segregation in exposure to political news on {Facebook},'' \emph{Science}, vol. 381, no. 6656, p. 392–398, July 2023.

\bibitem{allport1954prejudice}
G.~W. Allport, \emph{The nature of prejudice}.\hskip 1em plus 0.5em minus 0.4em\relax Cambridge, MA: Addison-Wesley, 1954.

\bibitem{Bai2020diversity}
X.~Bai, M.~R. Ramos, and S.~T. Fiske, ``As diversity increases, people paradoxically perceive social groups as more similar,'' \emph{Proc Natl Acad Sci}, vol. 117, no.~23, p. 12741–12749, May 2020.

\bibitem{degroot1974reaching}
M.~H. DeGroot, ``Reaching a consensus,'' \emph{J Am Stat Assoc}, vol.~69, no. 345, pp. 118--121, 1974.

\bibitem{friedkin1990social}
N.~E. Friedkin and E.~C. Johnsen, ``Social influence and opinions,'' \emph{J Math Sociol}, vol.~15, no. 3-4, pp. 193--206, 1990.

\bibitem{friedkin1999influence}
------, ``Social influence networks and opinion change,'' \emph{Adv Group Process}, vol.~16, no.~1, pp. 1--29, 1999.

\bibitem{proskurnikov2017tutorial}
A.~V. Proskurnikov and R.~Tempo, ``A tutorial on modeling and analysis of dynamic social networks. {P}art {I},'' \emph{Annu Rev Control}, vol.~43, pp. 65--79, 2017.

\bibitem{proskurnikov2018tutorial}
------, ``A tutorial on modeling and analysis of dynamic social networks. {P}art {II},'' \emph{Annu Rev Control}, vol.~45, pp. 166--190, 2018.

\bibitem{Deffuant2000}
G.~Deffuant, D.~Neau, F.~Amblard, and G.~Weisbuch, ``Mixing beliefs among interacting agents,'' \emph{Adv Complex Syst}, vol.~03, no. 01n04, pp. 87--98, Jan. 2000.

\bibitem{Lorenz}
J.~Lorenz, \emph{Fostering Consensus in Multidimensional Continuous Opinion Dynamics under Bounded Confidence}.\hskip 1em plus 0.5em minus 0.4em\relax Berlin Heidelberg: Springer, 2008, pp. 321--334.

\bibitem{Schweighofer2020}
S.~Schweighofer, D.~Garcia, and F.~Schweitzer, ``An agent-based model of multi-dimensional opinion dynamics and opinion alignment,'' \emph{Chaos}, vol.~30, no.~9, 2020.

\bibitem{Peng2021-cr}
{Peng et al.}, ``\BIBforeignlanguage{en}{A multilayer network model of the coevolution of the spread of a disease and competing opinions},'' \emph{\BIBforeignlanguage{en}{Math. Models Methods Appl. Sci.}}, vol.~31, no.~12, pp. 2455--2494, Nov. 2021.

\bibitem{Parsegov2017}
S.~E. Parsegov, A.~V. Proskurnikov, R.~Tempo, and N.~E. Friedkin, ``Novel multidimensional models of opinion dynamics in social networks,'' \emph{IEEE Trans Autom Control}, vol.~62, no.~5, pp. 2270--2285, May 2017.

\bibitem{Bizyaeva2023}
A.~Bizyaeva, A.~Franci, and N.~E. Leonard, ``Nonlinear opinion dynamics with tunable sensitivity,'' \emph{IEEE Trans Autom Control}, vol.~68, no.~3, p. 1415–1430, Mar. 2023.

\bibitem{Jia2015}
P.~Jia, A.~MirTabatabaei, N.~E. Friedkin, and F.~Bullo, ``Opinion dynamics and the evolution of social power in influence networks,'' \emph{SIAM REV}, vol.~57, no.~3, p. 367–397, Jan. 2015.

\bibitem{Watts1998}
D.~J. Watts and S.~H. Strogatz, ``\BIBforeignlanguage{en}{Collective dynamics of 'small-world' networks},'' \emph{\BIBforeignlanguage{en}{Nature}}, vol. 393, no. 6684, pp. 440--442, June 1998.

\end{thebibliography}

\end{document}